\documentclass[a4paper]{article}

\usepackage{graphicx}
\usepackage{dcolumn}
\usepackage{bm}
\usepackage{amsmath,amssymb}

\title{Extraordinary increase of lifetime of localized cold clouds
by the viscous effect in thermally-unstable two-phase interstellar media}
\author{Hiroki Yatou and Sadayoshi Toh \\
Department of Physics, Graduate School of Science, \\
Kyoto University, Kyoto 606-8502, Japan
}
\date{\today}

\begin{document}

\maketitle

\begin{abstract}
We numerically examine the influence of the viscosity on 
the relaxation process of localized clouds in thermally unstable 
two-phase media, which are locally heated by cosmic ray and cooled by radiation.
Pulselike stationary solutions of the media
are numerically obtained by a shooting method.
In one-dimensional direct numerical simulations, 
localized clouds are formed during the two-phase separation and
sustained extraordinarily.
Such  long-lived clouds have been recently observed in interstellar media.
We demonstrate that the balance of the viscosity with a pressure 
gradient remarkably suppresses the evaporation of the clouds and
controls the relaxation process. This balance fixes the peak pressure of
localized structures and then the structure is attracted and trapped to
one of the pulselike stationary solutions. 
While the viscosity has been neglected in most of previous studies,
our study suggests that the precise treatment of the
viscosity is necessary to discuss the evaporation of the clouds.
\end{abstract}

\section{Introduction}
Structures, especially spatially localized ones such as interfaces,
shocks, and pulses, play a crucial  role in understanding elementary 
processes, dynamical nature and even statistics  of widespread 
nonlinear phenomena 
(e.g., \cite{Toh:1987,Iwasaki_Toh:1992,Itano_Toh:2001,
Kawahara_Kida:2001,Pipe:2007,Waleffe:2003}).
In some cases, these structures are represented by
stationary or steady, even periodic solutions, e.g.
heteroclinic orbit that is a front connecting two states
and homoclinic orbit corresponding to a pulse.

In this paper, we try to apply this approach to 
the formation process of interstellar clouds that are cold, 
dense structures surrounded by warm phase in  thermally bistable media. 
This type of structures is often observed in nature: intergalactic clouds \cite{Davidson:1972},
and plasmas in tokamaks \cite{Lipschultz:1987}
(see also references in the work of Meerson \cite{Meerson:1996}).
These cold structures have attracted much attention in recent years in 
astrophysical contexts such as star formation processes 
(e.g., \cite{Audit:2005,Heitsch:2005,Vazquez:2006}).

Researches on interstellar clouds so far have focused on localized 
structures but have not systematically dealt with stationary solutions
and also their role in the formation process of clouds.
This is partially because such solutions are not stable and
do not survive in direct numerical simulations. However, 
unstable solutions can play a crucial role, e.g., traveling waves 
and unstable periodic orbits in turbulent production of channel 
flow turbulence \cite{Itano_Toh:2001,Kawahara_Kida:2001,Pipe:2007,Waleffe:2003}.
We, therefore, try to obtain homoclinic orbits that are  
spatially localized stationary solutions of the governing equations
and to describe the evolution of neutral and thermally bistable 
interstellar media by them.

Our main concern is the effect of viscosity on both formation of and 
saturation to localized structures, although the viscous effect 
has been  neglected or at most implicitly included as numerical 
viscosity because of its expected smallness in interstellar scales. 
In fact, in the studies so far, such structures are generally 
lost during the separation process \cite{Elphick:1992,Aranson:1993}.
However, as shown later, by including the viscous effect the evaporation 
time of the localized clouds is lengthened extraordinarily, where they 
are trapped close to stationary solutions. Physically speaking, 
the balance between  the viscosity and the pressure gradient  at a
higher order controls the saturation process. This might be related with 
recently observed tiny long-lived clouds \cite{Braun:2005,Stanimirovic:2005,Nagashima:2006}.

Here, we  will shortly review localized structures in interstellar
neutral media.
For homogeneous and stationary cases, radiative-equilibrium states are 
attained through the balance of radiative effects at each
point in space: heating due to cosmic ray  and radiative cooling \cite{Parker:1953}.
Field \textit{et al.} \cite{Field:1969}
found that three of such states exist for a range of pressure and only
two of them are linearly stable  with respect to spatial disturbances. 
If the system is initially in the unstable state, i.e., thermally
unstable one (\cite{Field:1965}), 
due to the spatial instabilities the system spontaneously separates into 
regions of stable states called the warm neutral media and the cold
neutral media.

For one-dimensional cases, cold structures are transiently formed 
during the separation process.  Some of them merge or evaporate, 
and finally few clouds survive. They are  surrounded by fronts
connecting two stable radiatively equilibrium states.
Zel'dovich and Pikel'ner (hereafter ZP) \cite{Zeldovich:1969}
studied the single-front dynamics in terms of steady solutions with 
one front in a plane-parallel geometry under an isobaric condition. 
The front is a traveling wave but stationary only when pressure
takes a certain value called saturation pressure.
This nature seems to resemble two-phase systems described 
by complex Ginzburg-Landau equations. However, it should be 
noted that even in the stationary state, the system is sustained 
by thermal flux compensating excesses of local energy budgets. 
In this sense, the system is essentially thermally nonequilibrium.
 
From the view of structures, most of the succeeding studies are 
based on fronts. Elphick \textit{et al.} \cite{Elphick:1992} extended 
the ZP solutions to a multifront case with an open boundary condition. 
In their numerical simulation, young cold clouds are merged and 
a uniform state is finally realized. However, rigid boundary 
conditions \cite{Aranson:1993} changed the destination of cold clouds. 
A single system-size cold cloud constituted by two fronts 
was sustained. The conservation of mass in the numerical domain  
governs the final state.

For two- or three-dimensional cases, localized structures with symmetries 
as well as plane-wave fronts have been investigated.
Graham and Langer \cite{Graham:1973} obtained
spherically symmetric steady solutions under an isobaric condition. 
The  expanding spherical front, i.e., condensing cold cloud,
asymptotically coincides with the ZP solution as the radius goes to infinity.
Nagashima \textit{et al.} \cite{Nagashima:2006} also calculated
the three-dimensional spherically symmetric flow
by solving the equations numerically with an open boundary condition.
They obtained the stationary spherical 
clouds only when the clouds have critical radii depending on
pressure, otherwise the curvature effect cannot balance the thermal effects. 
In order to compare the spherical clouds to
relatively long-lived tiny interstellar clouds observed recently,
they estimated the evaporation and condensation timescales  for contracting
(evaporating) and expanding (condensing) spherically-symmetric fronts
of the cold clouds respectively, and concluded that the estimated timescales
are consistent with the observations.

This paper is organized in the following way.
Section\ \ref{sec:equations} provides governing equations and some
important physical quantities.
Pulselike stationary solutions are numerically obtained in Sec.\ \ref{sec:localsolution}.
Section\ \ref{sec:localdynamics} shows that the solutions are confirmed in direct numerical simulations.
Section\ \ref{sec:pressure} describes the effects of the viscosity and pressure gradient in the dynamics of the localized clouds.
We present summary, discussion, and conclusion in Sec.\ \ref{sec:discussion}.

\section{Governing equations}
\label{sec:equations}

\begin{figure}
\begin{center}
\includegraphics[scale=0.7]{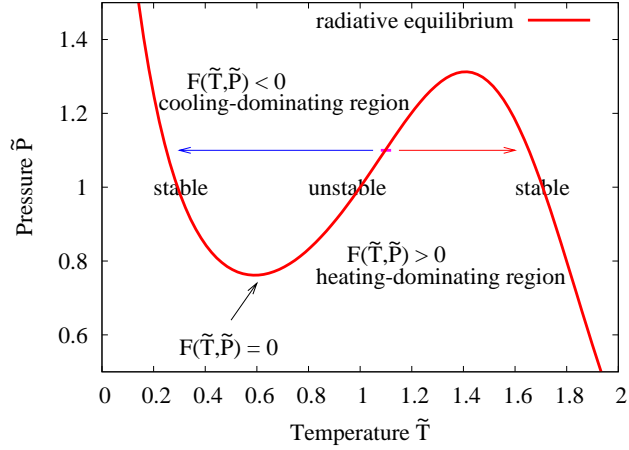}
\end{center}
\caption{
(Color online) Radiative-equilibrium curve (thick curve) of the
 Ginzburg-Landau-type heating-cooling function
(\ref{eq:ginz}) in the $\widetilde{T}\widetilde{P}$ plane.
The heating-cooling function $F$ equals zero on the curve.
$a$, $b$, and $c$ in Eq.\ (\ref{eq:ginz}) are set to $1.0$, $2.0$, and $1.0$, respectively.
The two arrows schematically denote that a small perturbation on an unstable state grows,
and one of the stable two-phase states with lower or higher temperature
on the radiative-equilibrium curve is finally realized.
}
\label{fig:tpgraph_schem}
\end{figure}

Consider an optically thin neutral media with external heating and radiative cooling.
Self-gravitation, magnetic field, and other body forces are neglected.
The fluids are described through the following basic equations which
include the laws of mass,
momentum, and energy conservation, together with a perfect-gas equation 
of state:
\begin{equation}
\frac{\partial \rho}{\partial t} + \frac{\partial}{\partial x_j}(\rho u_j) = 0,
\label{eq:origcon}
\end{equation}
\begin{equation}
\frac{\partial}{\partial t}(\rho u_i) + \frac{\partial}{\partial x_j}(\rho u_i u_j
-\sigma_{ij}) = -\frac{\partial P}{\partial x_i},
\end{equation}
\begin{equation}
\frac{\partial E}{\partial t} + \frac{\partial}{\partial x_j}\left[(E + P)u_j
- \sigma_{ij}u_i - K\frac{\partial T}{\partial x_j}\right] = -\rho {\cal L},
\end{equation}
\begin{equation}
\sigma_{ij} = \mu\left[\left(\frac{\partial u_i}{\partial
 x_j}+\frac{\partial u_j}{\partial x_i}\right)\
 -\frac{2}{3}\delta_{ij}\frac{\partial u_k}{\partial x_k}\right],
\end{equation}
\begin{equation}
P = \frac{\rho k_BT}{m_{\mathrm{H}}} = nk_BT,
\end{equation}
\begin{equation}
E \equiv \frac{P}{\gamma-1}+\frac{1}{2}\rho u_i^2,
\label{eq:origint}
\end{equation}
where $\rho$, $v$, $E$, and $P$ are density, velocity, total-energy density, and pressure, respectively.
The thermal conductivity, the specific-heat ratio, and the viscosity
coefficient are denoted by $K$, $\gamma$, and $\mu$, respectively.
In general the thermal conductivity is a function of
temperature but for simplicity we set it as a  constant.
$m_{\mathrm{H}}$ is the molecular mass of hydrogen and $k_B$ is the Boltzmann constant.
${\cal L}(T,P)$ is called a heating-cooling function, which describes the total of thermal
input and output depending on temperature and  pressure at each point.
Note that fluids can be locally heated and cooled in this system
since the fluids are sufficiently rarefied.  Such fluids are common particularly in astrophysical systems.

We use the following non-dimensional equations derived from the original
Eqs. (\ref{eq:origcon})-(\ref{eq:origint}):
\begin{equation}
\frac{\partial \widetilde{\rho}}{\partial \tau} + \frac{\partial}{\partial \widetilde{x}_j}
(\widetilde{\rho} \widetilde{u}_j) = 0,
\label{eq:continuum}
\end{equation}
\begin{equation}
\frac{\partial}{\partial \tau}(\widetilde{\rho}\widetilde{u}_i)+\frac{\partial}{\partial \widetilde{x}_j}
\left(\widetilde{\rho}\widetilde{u_i}\widetilde{u_j} -\frac{\gamma-1}{\gamma}\frac{\Pr}{E_p}
\widetilde{\sigma}_{ij}\right) = -\frac{\partial \widetilde{P}}{\partial \widetilde{x_i}},
\label{eq:kinetic}
\end{equation}
\begin{eqnarray}
Ep\left[\frac{1}{\gamma-1}\frac{\partial \widetilde{P}}{\partial \tau}
+ \frac{1}{\gamma-1}\frac{\partial}{\partial \widetilde{x}_j}(\widetilde{P}\widetilde{u}_j)
+ \widetilde{P}\frac{\partial \widetilde{u}_j}{\partial \widetilde{x}_j} \right] - \nonumber \\
\frac{\gamma-1}{\gamma}\Pr\widetilde{\sigma}_{ij}\frac{\partial \widetilde{u}_i}{\partial \widetilde{x}_j}
- \frac{\partial^2 \widetilde{T}}{\partial \widetilde{x}_j \partial \widetilde{x}_j}
= -\widetilde{\rho}\widetilde{\cal L},
\label{eq:energy}
\end{eqnarray}
\begin{equation}
\widetilde{P}=\widetilde{\rho}\widetilde{T},
\label{eq:state}
\end{equation}
\begin{equation}
\widetilde{\sigma}_{ij} = \left[\left(\frac{\partial \widetilde{u}_i}{\partial \widetilde{x}_j} +
\frac{\partial \widetilde{u}_j}{\partial \widetilde{x}_i}\right) -
\frac{2}{3}\delta_{ij}\frac{\partial \widetilde{u}_k}{\partial \widetilde{x}_k}\right].
\label{eq:stress}
\end{equation}
Here $\widetilde{x_j}=x_j/x_0$, $\widetilde{u_j}=u_j/u_0$, $\tau=t/(x_0/u_0)$,
$\widetilde{\rho}=\rho/\rho_0$, $\widetilde{T}=T/T_0$, $\widetilde{P}=P/P_0$,
and $\widetilde{{\cal L}}= {\cal L}/{\cal L}_0$,
which are scaled by the corresponding characteristic values denoted by a subscript zero, respectively.

The characteristic length scale $x_0$ is defined as follows:
\begin{equation}
 x_0 = \sqrt{\frac{KT_0}{\rho_0{\cal L}_0}}.
\end{equation}
This length is called Field length and represents the front width between two phases \cite{Field:1965}.

We set the characteristic velocity $u_0$ to be proportional to a sound speed:
\begin{equation}
 u_0 = \sqrt{\frac{k_B}{m_{\mathrm{H}}}T_0}.
\end{equation}

There are two non-dimensional constants in the non-dimensional equations:
\begin{equation}
\Pr=\frac{\gamma}{\gamma-1}\frac{k_B\mu}{m_{\mathrm{H}}K},
\end{equation}
\begin{equation}
Ep=\frac{1}{\rho_0{\cal L}_0}\frac{P_0}{t_0}.
\end{equation}
$\Pr$ is called Prandtl number.
$Ep$ represents the degree of thermal contribution to
dynamics in Eq.\ (\ref{eq:energy}). Larger $Ep$ induces slower dynamics in fluids.

\begin{figure}
\begin{center}
\includegraphics[scale=0.65,clip]{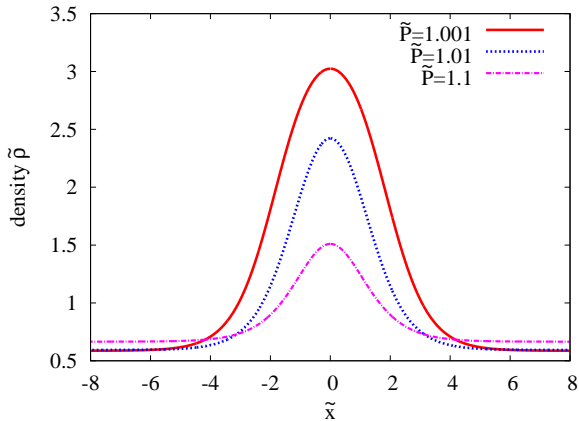}
\end{center}
\caption{
(Color online) Density profiles of pulselike stationary solutions
for the three values of pressure $\widetilde{P}=1.1$, $1.01$, and $1.001$.
}
\label{fig:solution_1d}
\end{figure}

\begin{figure}
\begin{center}
\includegraphics[scale=0.7]{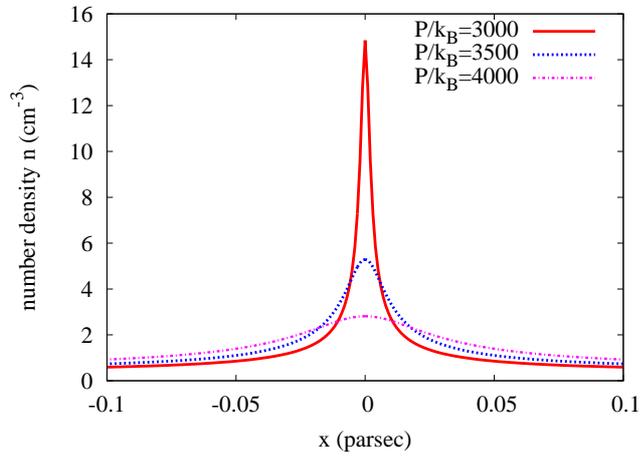}
\end{center}
\caption{
Pulselike stationary solutions obtained using the astrophysically realistic but complicated form
(\ref{eq:realform1}).}
\label{fig:pulse1d_astro}
\end{figure}

In this paper, the following Ginzburg-Landau-type equation is adopted
for the heating-cooling function \cite{Elphick:1991,Elphick:1992,Nagashima:2005}:
\begin{equation}
-\widetilde{\rho}\widetilde{\cal L}(\widetilde{T},\widetilde{P}) =
 F(\widetilde{T},\widetilde{P}) = a(\widetilde{T}-1)-b(\widetilde{T}-1)^3-c\ln\widetilde{P}.
\label{eq:ginz}
\end{equation}
We also tried the more realistic but complicated form of the
heating-cooling function (\ref{eq:realform1}) used in astrophysical contexts \cite{Koyama:2000,Koyama:2002}:
\begin{equation}
\rho {\cal L}=-n\Gamma+n^2\Lambda(T),
\label{eq:realform1}
\end{equation}
\begin{equation}
\Gamma=2.0\times10^{-26}\ \mathrm{ergs}\ \mathrm{s}^{-1},
\label{eq:realform2}
\end{equation}
\begin{eqnarray}
\frac{\Lambda(T)}{\Gamma} = & 1.0\times10^7\exp [-1.184\times10^5/(T+1000)] \nonumber \\ 
+ & 1.4\times10^{-2}\sqrt{T}\exp(-92/T)\ \mathrm{cm}^3.
\label{eq:realform3}
\end{eqnarray}
It is found that both the stationary solutions and the results of numerical simulations 
with the heating-cooling function (\ref{eq:realform1})
are qualitatively similar to those with the simplified heating-cooling function (\ref{eq:ginz})
 (see Figs.\ \ref{fig:solution_1d},\ref{fig:pulse1d_astro} and
Figs.\ \ref{fig:evldensity_rand},\ref{fig:final1d_astro}).
Therefore, for simplicity and for elucidation of fundamental mechanism,
the simplest form of the heating-cooling function defined by
Eq.\ (\ref{eq:ginz}) is adopted in this work.

In Fig.\ \ref{fig:tpgraph_schem}, we show a radiative equilibrium curve
on which the heating-cooling function $F(\widetilde{T},\widetilde{P})$ is equal to zero.
The S shape of the curve is essential to two-phase separation:
supposed to a fixed pressure in a range with which the equation has three roots.
Because two roots with high and low temperatures among them satisfy condition\ (\ref{eq:stable}),
they are stable against small temperature (density) perturbations,
\begin{equation}
\left(\frac{\partial F}{\partial \widetilde{T}}\right)_{\widetilde{P}} < 0.
\label{eq:stable}
\end{equation}
In contrast, the root with an intermediate temperature satisfies condition
(\ref{eq:unstable}), and thus a small disturbance on the state grows.
This instability is called thermal instability \cite{Field:1965},
\begin{equation}
\left(\frac{\partial F}{\partial \widetilde{T}}\right)_{\widetilde{P}} > 0.
\label{eq:unstable}
\end{equation}
Fluid particles around the unstable branch in Fig.\ \ref{fig:tpgraph_schem} finally
reach to each of stable branches with high or low temperatures via heating or cooling processes.

\section{Pulselike stationary solutions}
\label{sec:localsolution}

\begin{figure}
\begin{center}
\includegraphics[scale=0.65,clip]{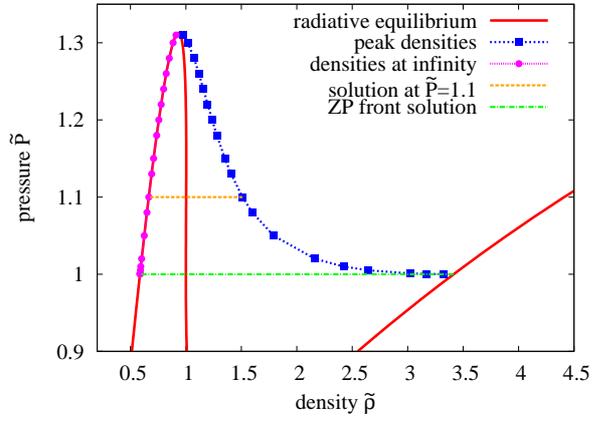}
\end{center}
\caption{
(Color online) Peak densities $\widetilde{\rho}_0$ and densities at infinity $\widetilde{\rho}_{\infty}$
of the pulselike stationary solutions.
The dashed horizontal lines denote the projections of the solution with $\widetilde{P}=1.1$
and the ZP front solution with the saturation pressure $\widetilde{P}=1$.
}
\label{fig:tpgraph_nondim_1d}
\end{figure}

\begin{figure*}
\begin{center}
\includegraphics[scale=0.65,clip]{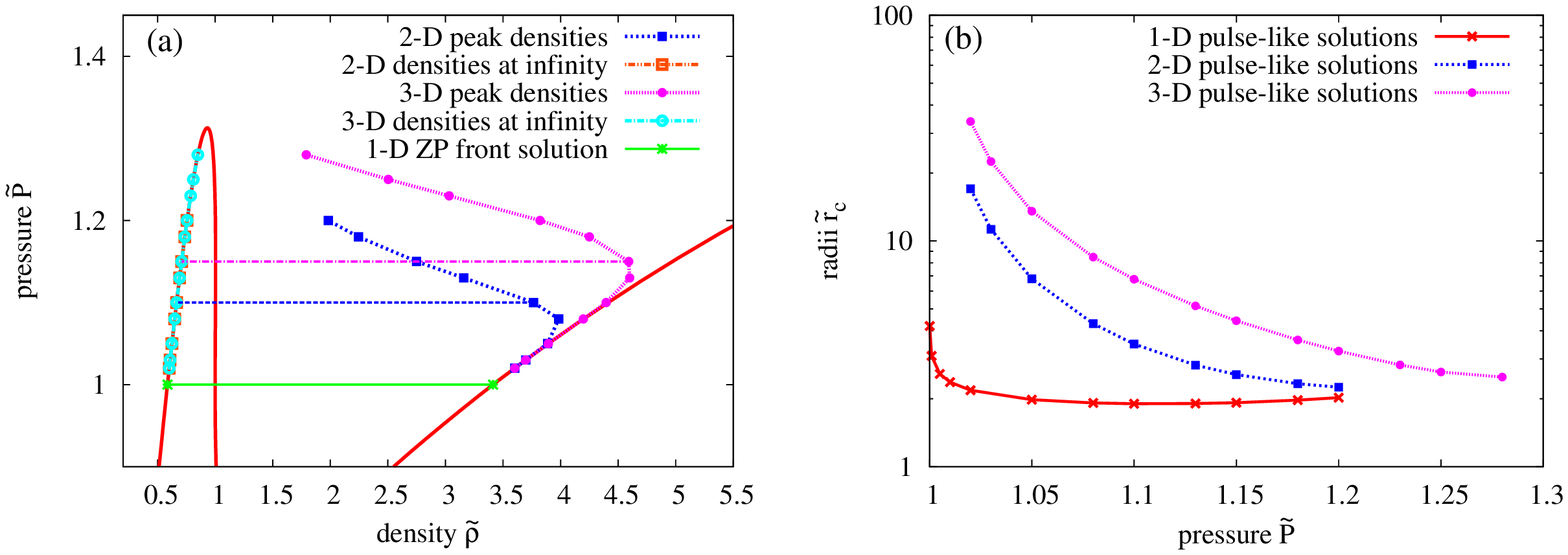}
\end{center}
\caption{
(a) Peak densities $\widetilde{\rho}_0$ and densities at infinity $\widetilde{\rho}_{\infty}$ 
of the pulselike stationary solutions 
of Eqs.\ (\ref{eq:multidim})$-$(\ref{eq:boundarym}) in two-dimensional and three-dimensional cases,
respectively.
The two-dimensional and three-dimensional densities at infinity almost overlap.
The two dashed horizontal lines with $\widetilde{P} > 1$ represent the projections of the density profiles
of some two-dimensional and three-dimensional solutions into the $\widetilde{\rho}\widetilde{P}$ plane.
The horizontal line with $\widetilde{P}=1$ represents the one-dimensional ZP front solution.
(b) Radii of the pulselike stationary solutions as a function of pressure $\widetilde{P}$
in one-dimensional, two-dimensional axisymmetric, and three-dimensional spherically symmetric systems.
}
\label{fig:tpgraph_nondim_2d3d}
\end{figure*}

First, we will obtain stationary localized solutions of Eqs.\
(\ref{eq:continuum})$-$(\ref{eq:stress}).
These are homoclinic orbits.

\subsection{Plane-parallel geometry}

Supposed that a fluid is 
stationary ($\partial/\partial \tau = 0$ and $\widetilde{u}_i=0$),
the governing equations are simplified 
in a plane-parallel geometry as follows:
\begin{equation}
F(\widetilde{T},\widetilde{P}) +
\frac{\partial^2 \widetilde{T}}{\partial \widetilde{x}^2} = 0,
\label{eq:steadystatic}
\end{equation}
\begin{equation}
\frac{\partial \widetilde{P}}{\partial \widetilde{x}} = 0.
\label{eq:steadypressure}
\end{equation}
Because the pressure $\widetilde{P}$ becomes uniform via Eq.\ (\ref{eq:steadypressure}),
the pressure works as a parameter in the temperature Eq.\ (\ref{eq:steadystatic}).
Here for simplicity and for keeping the temperature positive, 
 we set the parameters $a$, $b$, and $c$ of $F$ as 1, 2 and 1,
respectively. However, note that transforming the
variables as $\widetilde{x}=\sqrt{a}\xi$, $\widetilde{T}-1=\sqrt{a/b}\hat{T}$, and
$\widetilde{P}=\hat{P}^{\sqrt{ab}/c}$, the parameters can be eliminated from
Eq.\ (\ref{eq:steadystatic}).

We impose the following boundary conditions in the semi-infinite space ($0 \leq \widetilde{x} < \infty$):
\begin{equation}
\frac{\partial \widetilde{T}}{\partial \widetilde{x}} = 0
\hspace{20pt} (\widetilde{x} = 0,\ \widetilde{x} \to \infty).
\label{eq:boundary}
\end{equation}
Since Eq.\ (\ref{eq:steadystatic}) is a second-order ordinary differential equation
with respect to $\widetilde{x}$, the two boundary conditions in Eq.\ (\ref{eq:boundary})
uniquely determine the solution $\widetilde{T}(\widetilde{x})$ if it exists.
Simultaneously, the values of temperature at the boundaries, i.e., 
$\widetilde{T}_0=\widetilde{T}(\widetilde{x}=0)$ and 
$\widetilde{T}_{\infty}=\widetilde{T}(\widetilde{x}\to \infty)$, 
are given.
We numerically solve Eq.\ (\ref{eq:steadystatic}) with 
the boundary conditions (\ref{eq:boundary})
by means of a shooting method, and we obtain $\widetilde{T}(\widetilde{x})$ and
also $\widetilde{T}_0$ and $\widetilde{T}_{\infty}$
for various values of pressure.
In Fig.\ \ref{fig:solution_1d}, we show the density profiles 
$\widetilde{\rho}(\widetilde{x}) = \widetilde{P}/\widetilde{T}(\widetilde{x})$
of the solutions for three values of pressure $\widetilde{P}=1.1$, $1.01$, and $1.001$.
By the reflection symmetry of Eq.\ (\ref{eq:steadystatic})
the solutions are extended to $\widetilde{x}<0$.
These localized cold solutions are called pulselike stationary solutions, hereafter.
Figure\ \ref{fig:tpgraph_nondim_1d} shows the peak density
 $\widetilde{\rho}_0=\widetilde{P}/\widetilde{T}_0$
and the density at infinity
 $\widetilde{\rho}_{\infty}=\widetilde{P}/\widetilde{T}_{\infty}$ 
as a function of pressure $\widetilde{P}$.
Note that the peak densities are
always lower than the density of the stable equilibrium state and
 apart from the radiative-equilibrium curve ($F=0$) when $\widetilde{P} > 1$.
As $\widetilde{P} \to 1+$, the peak density
asymptotically approaches one of the stable radiative equilibriums, 
and the pulselike solution becomes the ZP solution (front solution).

We also obtained pulselike stationary solutions using the realistic
but complicated form of heating-cooling function.
The solutions are shown in Fig.\ \ref{fig:pulse1d_astro}.
With the respect that the solutions are also homoclinic orbits,
they are qualitatively similar to those obtained with simplified heating-cooling function (\ref{eq:ginz})
as shown in Fig.\ \ref{fig:solution_1d}.
This supports the use of the simplified heating-cooling function\ (\ref{eq:ginz}).

\subsection{Axisymmetric and spherically-symmetric geometry}
\label{sec:axis}

\begin{figure*}
\begin{center}
\includegraphics[scale=0.65,clip]{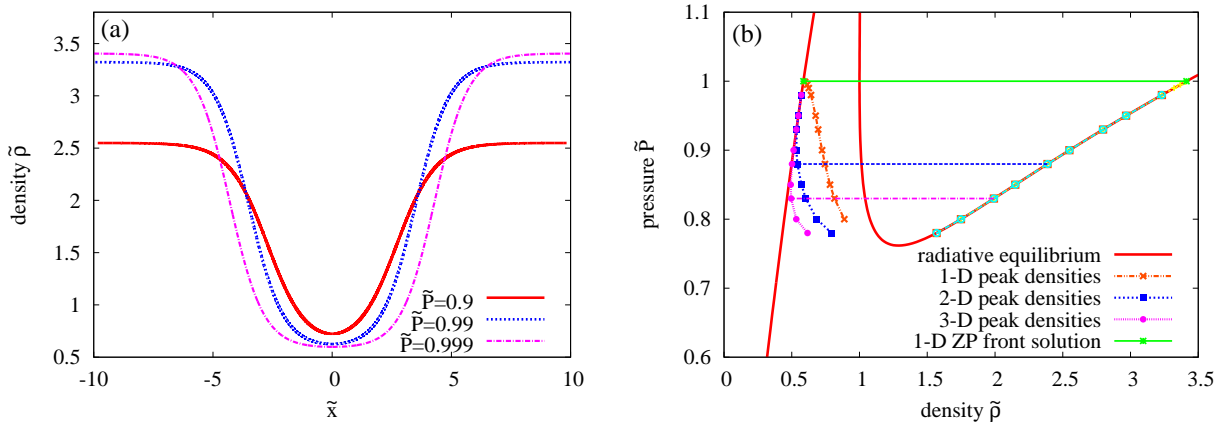}
\end{center}
\caption{
(Color online) (a) Density profiles of the warm pulselike stationary solutions of 
Eqs.\ (\ref{eq:steadystatic}) and (\ref{eq:boundary})
for the three values of pressure $\widetilde{P}=0.9$, $0.99$, and $0.999$.
(b) Peak densities of the solutions in one-dimensional, two-dimensional axisymmetric,
and three-dimensional spherically symmetric systems in the $\widetilde{\rho}\widetilde{P}$ plane.
The two horizontal lines with $\widetilde{P} < 1$ represent the projections of the density profiles
of some two-dimensional and three-dimensional solutions into the $\widetilde{\rho}\widetilde{P}$ plane.
The horizontal line with $\widetilde{P}=1$ represents the one-dimensional ZP front solution.
}
\label{fig:bubble}
\end{figure*}

We also obtain pulselike stationary solutions 
in multidimensional systems as in the one-dimensional case.
We suppose that the systems are axisymmetric in two-dimensional cases
and spherically symmetric in three-dimensional cases.
These symmetries make the governing equations simpler as follows:
\begin{eqnarray}
F(\widetilde{T},\widetilde{P})
+ \frac{d-1}{\widetilde{r}}\frac{\partial \widetilde{T}}{\partial \widetilde{r}}
+ \frac{\partial^2 \widetilde{T}}{\partial \widetilde{r}^2} &=& 0, \label{eq:multidim} \\
\frac{\partial \widetilde{P}}{\partial \widetilde{r}} &=&  0,
\end{eqnarray}
where $\widetilde{r}$ and $d$ are the radial coordinate and 
 the space dimension, respectively.
The boundary conditions are imposed as follows:
\begin{equation}
\frac{\partial \widetilde{T}}{\partial \widetilde{r}} = 0
\hspace{20pt} (\widetilde{r} = 0, \widetilde{r} \to \infty).
\label{eq:boundarym}
\end{equation}

These equations are numerically solved by means of the shooting method.
As in the case of the one-dimensional system, we obtain a localized solution 
$\widetilde{\rho}(\widetilde{r}) =\widetilde{P}/\widetilde{T}(\widetilde{r})$ 
with peak density $\widetilde{\rho}_0$ at the origin and density at infinity
 $\widetilde{\rho}_{\infty}$ at $\widetilde{r} \to \infty$.
The two densities are dependent only on pressure.
Figure\ \ref{fig:tpgraph_nondim_2d3d}(a) shows $\widetilde{\rho}_0$ and $\widetilde{\rho}_{\infty}$
as a function of pressure $\widetilde{P}$ for the two-dimensional and three-dimensional solutions.
 As pressure increases from the saturation pressure $\widetilde{P}=1$,
the peak density leaves away from the radiative-equilibrium curve.
Note that by the curvature effect represented by the second term in the left-hand side
of Eq.\ (\ref{eq:multidim}), the peak density increases
along with the radiative-equilibrium curve as pressure increases,
unlike the one-dimensional solution in Fig.\ \ref{fig:tpgraph_nondim_1d}.

We define the size of the localized structures as a radius $\widetilde{r}_c$,
which satisfies $\widetilde{T}(\widetilde{r}_c)=(\widetilde{T}_0+\widetilde{T}_{\infty})/2$.
Figure\ \ref{fig:tpgraph_nondim_2d3d}(b) shows the radii as a function of pressure.
When $\widetilde{P} \to 1+$, the radii go to infinity.
Since the curvature term of Eq.\ (\ref{eq:multidim}) asymptotically approaches zero in the limit,
the localized solutions then lead to the ZP front solution in a plane-parallel geometry.
The dependence of the cloud size on pressure qualitatively agrees
 with the results of Nagashima \textit{et al.} \cite{Nagashima:2006}.

\subsection{Bubble solutions}

Warm, rarefied, and localized structures surrounded by cold dense media
are also numerically obtained 
in the same way as the cold structures obtained in Sec.\ \ref{sec:axis}.
The density profiles of the solutions
for the three values of pressure $\widetilde{P}=0.9$, $0.99$, and $0.999$
are shown in Fig.\ \ref{fig:bubble}(a).
These structures are realized only when $\widetilde{P} < 1$.
Figure\ \ref{fig:bubble}(b) shows the peak densities $\widetilde{\rho}_0$
and densities at infinity $\widetilde{\rho}_{\infty}$  as a function of pressure $\widetilde{P}$.
As in the case of the cold structures, the peak densities are apart from the radiative-equilibrium
curve when  pressure gets smaller than the saturation pressure $\widetilde{P}=1$.
In the limit that $\widetilde{P} \to 1-$, the solutions approach the
one-dimensional ZP front solution corresponding to a 
plane-parallel and localized warm structure of an infinity size.
These localized warm solutions are called "bubbles" \cite{Aranson:1993,Meerson:1996}
and have been frequently referred in astrophysical contexts \cite{Meerson:1996}.

\section{relaxation process of localized cold structures}
\label{sec:localdynamics}

\begin{figure}
\begin{center}
\includegraphics[scale=0.65,clip]{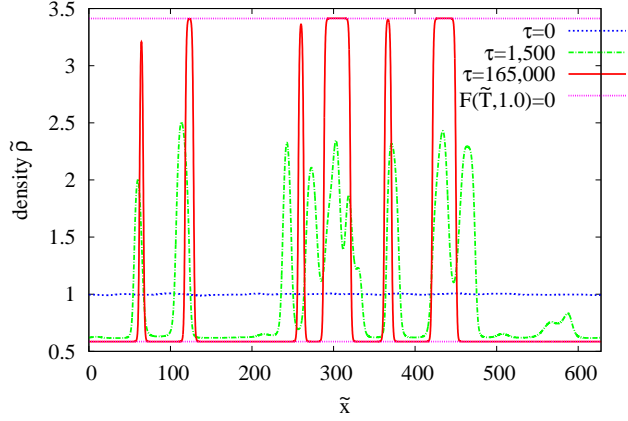}
\end{center}
\caption{
(Color online) The two-phase separation from the random small density disturbance
around $\widetilde{\rho} = 1$.
The density profiles at  $\tau=0$, $1500$ and $165 000$ are shown.
The times correspond to the initial, transient, and quasistationary state, 
respectively. The two horizontal lines represent the densities of 
the states of the radiative equilibrium with the saturation pressure $\widetilde{P}=1$.
}
\label{fig:evldensity_rand}
\end{figure}

\begin{figure}
\begin{center}
\includegraphics[scale=0.7]{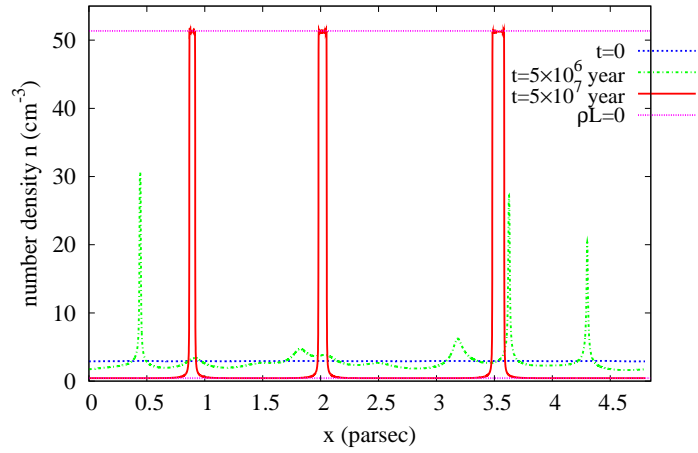}
\end{center}
\caption{
(Color online) Formation process of localized structures in direct numerical simulation
with the astrophysically realistic heating-cooling function (\ref{eq:realform1}).
The density profiles at the three times $t=0$, $5.0\times10^6 \;\mathrm{year}$, and
$5.0\times10^7 \;\mathrm{year}$ are plotted.
The two horizontal lines represent the densities of the states of the radiative equilibrium
at the saturation pressure.}
\label{fig:final1d_astro}
\end{figure}
\begin{figure}
\begin{center}
\includegraphics[scale=0.65]{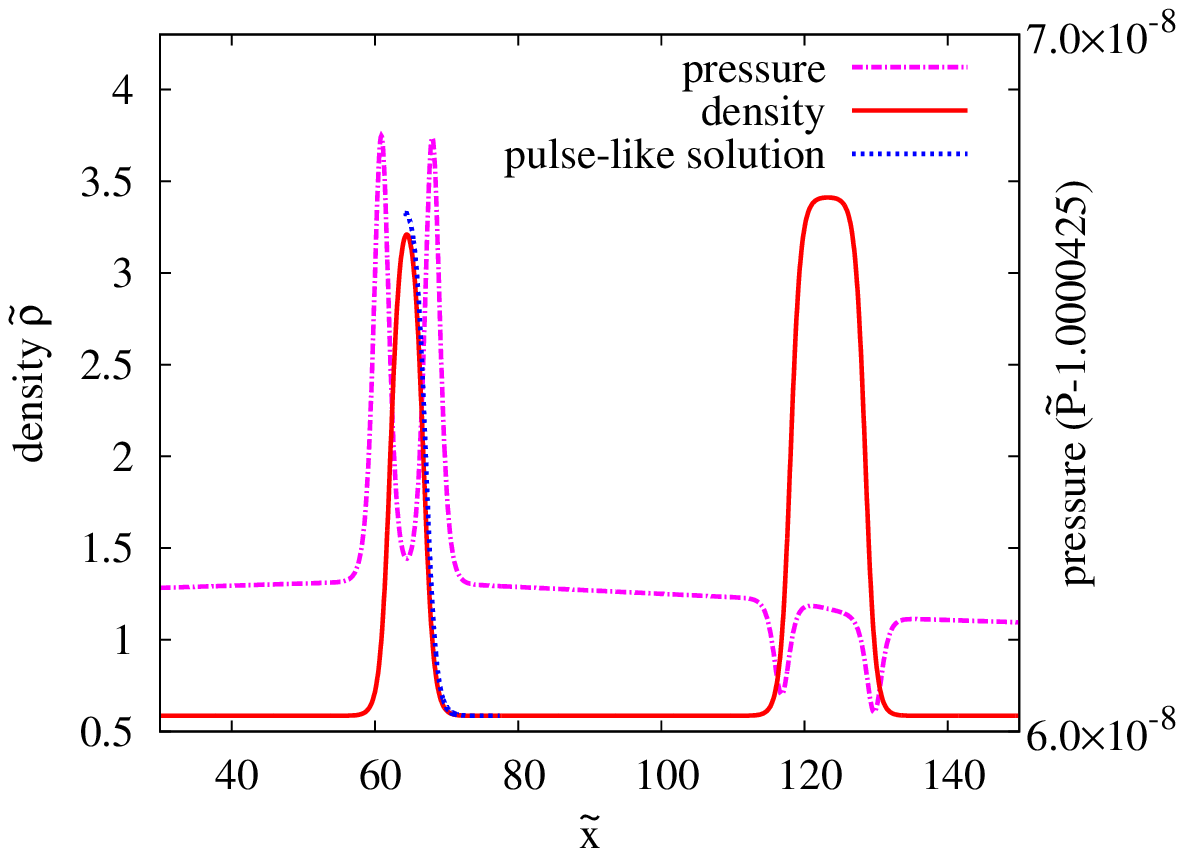}
\end{center}
\caption{
(Color online) Density and pressure profiles at $\tau=165 000$ in the quasistationary state. 
The pulselike stationary solution with the pressure value
at the coordinate where the localized structure has a peak density is superposed.
}
\label{fig:denspressure}
\end{figure}

We investigate the role of the cold pulselike stationary solutions
in the formation of localized structures by direct numerical simulations.
Equations\ (\ref{eq:continuum})-(\ref{eq:stress}) are numerically solved
in a plane-parallel geometry with 8192 grids and a periodic boundary condition.
We use a high-accuracy pseudospectral method partially 
dealiased and a fourth-order Runge-Kutta method.
As initial conditions, pressure is set to the saturation pressure 
and velocity is zero, that is, $\widetilde{P}=1$ and $\widetilde{u}=0$.
Density is a small random disturbance consisting of only 
low wave-number modes ($k \le 16$) added to a homogeneous state
 $\widetilde{\rho}=1$ that is on the unstable branch of the radiative-equilibrium
curve (see Fig.\ \ref{fig:tpgraph_schem}).
We set non-dimensional constants $Ep$ and $\Pr$ to
$1.0\times10^2$ and $1.0$, respectively. Koyama and Inutsuka
\footnote{H. Koyama and S. Inutsuka, e-print arXiv:astro-ph/0605528.}
used $Ep\simeq 64$ in a realistic situation so that 
$Ep=100$ seems to be a feasible value.
The specific-heat ratio $\gamma$ is set to 5/3.

Figure\ \ref{fig:evldensity_rand} shows a formation process of localized structures.
The initial density disturbance rapidly grows and evolves into warm
and cold regions due to a thermal instability.
Some of the cold regions merge successively and some of them evaporate in the evolution.
Several localized cold structures finally survive in warm media,
and then these structures persist for a long time.
Because most of the kinetic energy in the whole system is 
lost during the two-phase separation,
the cold structures are almost stationary.

For comparison, we also have numerically calculated the two-phase separation
with the realistic but complicated form of heating-cooling function
(\ref{eq:realform1}) and original astrophysical Eqs.\ (\ref{eq:origcon})$-$(\ref{eq:origint}).
Some of the results are shown in Fig.\ \ref{fig:final1d_astro}.
The relaxation process is qualitatively similar to that
with simplified heating-cooling function (\ref{eq:ginz}) shown in Fig.\ \ref{fig:evldensity_rand}.
Some of the cold regions merge and evaporate, and
some localized structures finally survive in the process.

To examine the decaying process of localized structures in detail,
we focus on two typical localized cold structures.  One with a longer lifetime
is seen around $x \sim 60$ in Fig.\ \ref{fig:evldensity_rand}. 
The peak density of this localized structure does not yet attain the 
value of the density in the radiative equilibrium ($F=0$).
However, as seen in Fig.\ \ref{fig:denspressure},
the localized cold structure is quite similar in form to a pulselike stationary solution 
 with the value of pressure at the peak of the localized structure.
The values of pressure  at the peak of localized
structures are called peak pressure.
Note that in Fig.\ \ref{fig:denspressure}  the value of pressure
is subtracted by 1.000 042 5 so that the pressure variation around 
the peak of the localized structure is negligible and does not 
affect the estimation of the peak pressure.

\begin{figure}
\begin{center}
\includegraphics[scale=0.65,clip]{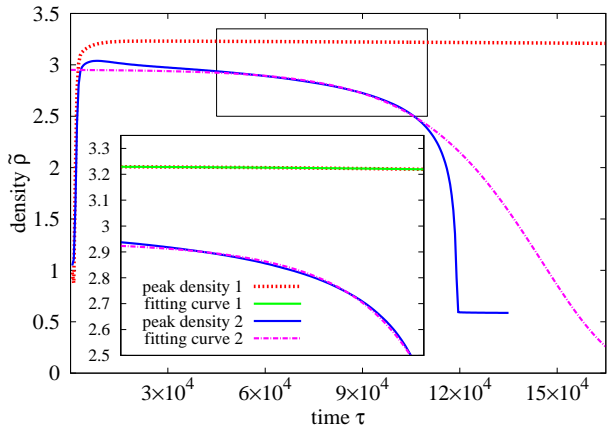}
\end{center}
\caption{
(Color online) Peak density evolutions of the localized structure
at $\widetilde{x} \sim 60$ in Fig.\ \ref{fig:evldensity_rand}
and the other localized structure derived by another run.
Inset: the peak density evolutions magnified in the 
quasistationary state $4.5\times 10^4 \leq \tau \leq 1.1\times 10^5$.
The curve with the larger peak density is fitted by the double-exponent function defined
as $A\exp\{-B\exp[(\tau-\tau_0)/\tau_c]\}$ with $A = 3.27$, $B = 1.24\times10^{-2}$,
and the relaxation time $\tau_c = 2.91\times10^5$.
The curve with the smaller peak density is fitted by the above double-exponent function
with $A = 2.95$, $B = 1.07\times10^{-2}$, and $\tau_c = 2.21\times10^4$.
}
\label{fig:peakdensity}
\end{figure}

The peak density of the localized structures decreases slowly and monotonically
by evaporation and they finally disappear. 
We estimate the characteristic time of the decaying process, i.e.,
the evaporation time of the localized structures.
In Fig.\ \ref{fig:peakdensity}, we show 
the evolution of the peak density of the localized structures
obtained in two runs under the same conditions except for the initial condition:
one with a longer lifetime is called LS1 and the other LS2 hereafter.
The evolution is separated into two periods: the fast transient one and the slow decay one.
We call the latter a quasistationary state.
The transient period finishes at around $\tau = 2.0\times10^3$.
This time corresponds to a sound crossing time defined as a time during which 
a sound wave passes  through the system.

The decay curves  of the peak density after attaining the
quasistationary state are well fitted by a double-exponential function 
$A\exp\{-B\exp [(\tau-\tau_0)/\tau_c]\}$ as shown
in the inset of Fig.\ \ref{fig:peakdensity};  
the values of the fitting parameters of LS1 are $A = 3.27$, $B = 1.24 \times 10^{-2}$,
$\tau_0 = 45 000$, and $\tau_c = 2.91\times10^5$ for $ 4.5\times 10^{4} \le \tau \le 1.6 \times 10^{5}$;  
those of LS2 for $ 4.5\times 10^{4} \le \tau \le 1.1\times 10^{5}$
are $A = 2.95$, $B = 1.07\times 10^{-2}$, $\tau_0 = 45 000$, and $\tau_c = 2.21\times10^4$. 
The latter case shows that  the late decaying process is  faster than 
a double-exponential function, but the characteristic time estimated 
is shorter than the period of the saturation process
$1.2\times 10^{5}$.
 In this sense, the characteristic time gives
a good estimate of the evaporation time of the localized structure.

It should be noted that the relaxation times $\tau_c$ are much longer than the sound crossing time.
In Sec.\ \ref{sec:pressure}, we discuss the details of this decaying process of the peak density.

\section{persistence of pulselike clouds with viscosity}
\label{sec:pressure}

\begin{figure}
\begin{center}
\includegraphics[scale=0.65]{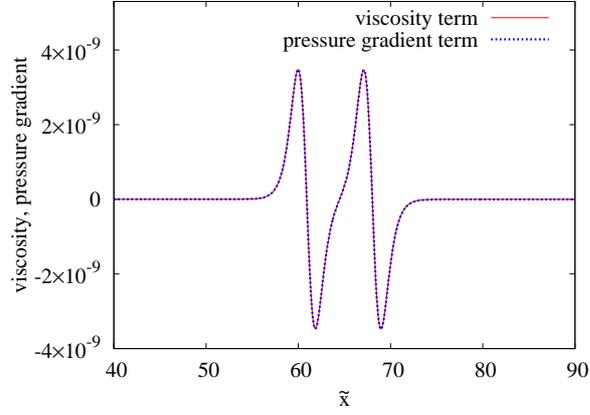}
\end{center}
\caption{
(Color online) The viscosity term and the pressure-gradient term of the kinetic Eq.\ (\ref{eq:kinetic})
in the quasistationary state with Prandtl number $\Pr = 1.0$ for LS1 at $\tau=165 000$.
}
\label{fig:previsc}
\end{figure}

\begin{figure}
\begin{center}
\includegraphics[scale=0.65]{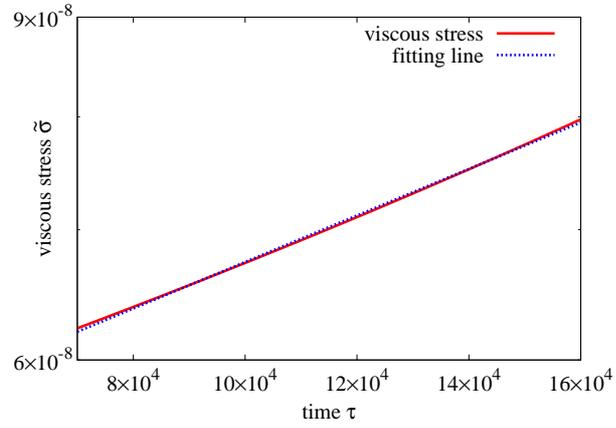}
\end{center}
\caption{
(Color online) The time evolution of the viscous stress $\widetilde{\sigma}$.
The curve is fitted by an exponential function: $\sigma_0\exp((\tau-\tau_0)/\tau_{c\sigma})$,
where $\tau_0 = 70 000$, $\sigma_0 = 6.20\times10^{-8}$, and $\tau_{c\sigma} = 3.63\times10^{5}$.
}
\label{fig:viscous_stress}
\end{figure}

\begin{figure*}
\begin{center}
\includegraphics[scale=0.65]{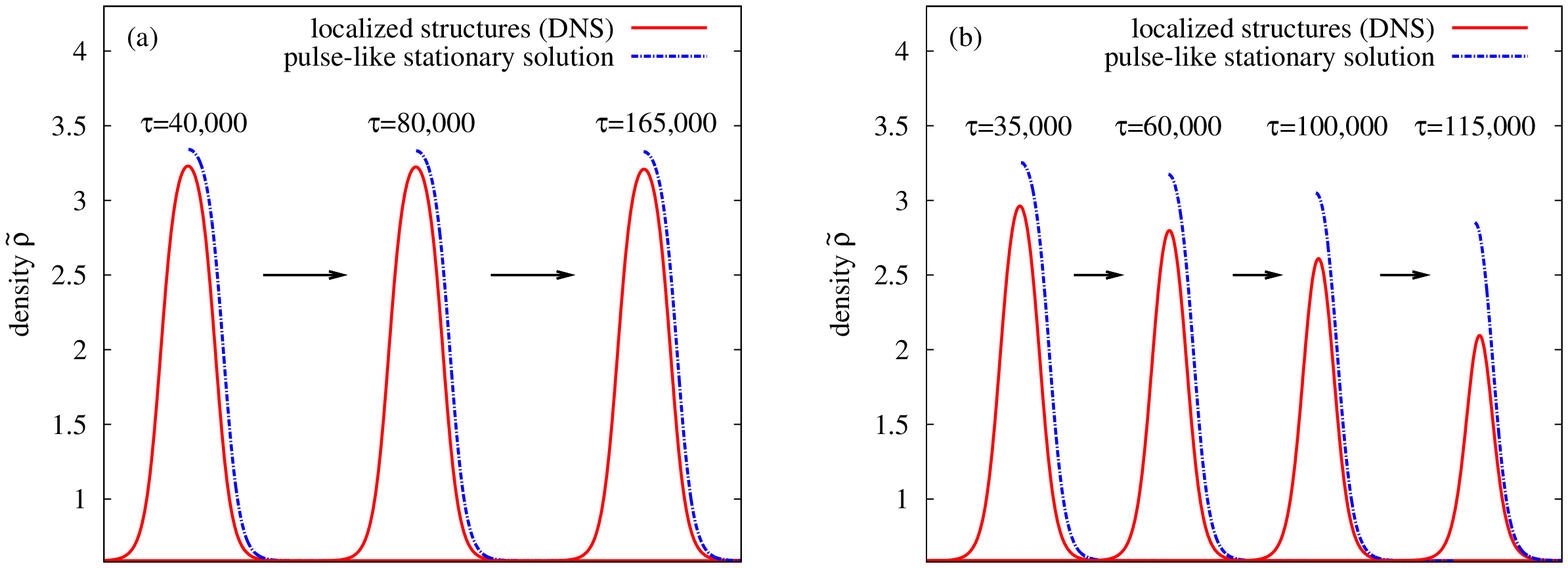}
\end{center}
\caption{
(Color online) Comparison of localized structures and corresponding
pulselike stationary solutions
in the relaxation process shown in Fig.\ \ref{fig:peakdensity}:
(a) LS1 and (b) LS2. The value of pressure of the  pulselike  solutions  
is set to the peak pressure of the localized structure at each time.
}
\label{fig:denssuccess}
\end{figure*}
Why can the localized cold structures be sustained extremely longer than expected?
Here we focus on viscosity, although the viscosity has been omitted in most of the previous studies. 
Since we have performed numerical simulations with a high-accuracy spectral scheme,
we can deal with the viscosity more precisely than the previous studies.

Figure\ \ref{fig:previsc} shows the viscosity term and the pressure-gradient term
of the kinetic equation (\ref{eq:kinetic}) around the localized structure
in the quasistationary state $\tau=165 000$.
While the pressure is almost uniform because of the mixing by sound waves,
a weak pressure gradient is maintained around localized structures
(see also Fig.\ \ref{fig:denspressure}).
Moreover,  the weak pressure gradient almost coincides with the viscosity.
This balance between the viscosity and the pressure gradient causes 
the persistence of the localized structure.
To see this in detail, we rewrite the basic equations (\ref{eq:continuum}) and
(\ref{eq:kinetic})
in a plane-parallel geometry as follows:
\begin{equation}
\frac{D\widetilde{\rho}}{D\tau} = -\frac{3}{4}\widetilde{\sigma}\widetilde{\rho},
\label{eq:continuum2}
\end{equation}
\begin{equation}
 \frac{D\widetilde{\sigma}}{D\tau}+\frac{3}{4}\widetilde{\sigma}^2 =
 \left( \frac{4}{3}\frac{1}{\widetilde{\rho}}\frac{\partial}{\partial \widetilde{x}}
 - \frac{4}{3}\frac{1}{\widetilde{\rho}^2}\frac{\partial \widetilde{\rho}}{\partial \widetilde{x}}\right)
 \left(\frac{\gamma-1}{\gamma}\frac{\Pr}{Ep}\frac{\partial\widetilde{\sigma}}
 {\partial \widetilde{x}}-\frac{\partial \widetilde{P}}{\partial \widetilde{x}}\right),
\label{eq:kinetic2}
\end{equation}
where $D/D\tau = \partial/\partial \tau + \widetilde{v}\partial/\partial \widetilde{x}$
is a Lagrangian derivative.
$\widetilde{\sigma} = 4/3(\partial \widetilde{v}/\partial \widetilde{x})$
is a viscous stress, which represents the degree of compressibility.

Figure\ \ref{fig:viscous_stress} shows that the viscous stress $\widetilde{\sigma}$
exponentially grows in the quasistationary state.
Hence, the Lagrangian time derivative $D\widetilde{\sigma}/D\tau$ can be written as
$\alpha\widetilde{\sigma}$, where the growth rate is $\alpha\sim10^{-6}$.
The very small growth rate is just caused by the small difference
between the viscosity and the pressure gradient in Eq.\ (\ref{eq:kinetic2})
when the nonlinear term in the left-hand side is negligible.
In fact $\widetilde{\sigma}\gg\alpha\widetilde{\sigma}\gg\widetilde{\sigma}^2$
since the viscous stress $\widetilde{\sigma}$ is  the order of $10^{-8}$
 as seen in Fig.\ \ref{fig:viscous_stress}.
The slow exponential growth of the viscous stress finally leads to
the double-exponential decay with a long relaxation time via Eq.\ (\ref{eq:continuum2}).

In this relaxation process, as shown in Fig.\ \ref{fig:denssuccess},
the localized structures remain close to pulselike stationary solutions.
The balance fixes the peak pressure of the localized structure.
Then the structure is attracted and trapped to one of the 
pulselike stationary solutions with the corresponding pressure value 
that seems to behave similar to a saddlelike fixed point in the phase space.
The deviation of the localized structure from the pulselike stationary
solution induces both the small pressure gradient and viscous effect (see
Fig.\ \ref{fig:previsc}). 
Finally the two effects induced balance to each other and then
this balance suppresses the induction of flow that transfers heat and 
material from localized structures.

The one-dimensional Eqs.\ (\ref{eq:continuum2}) and (\ref{eq:kinetic2}) can be extended to
multidimensional equations as follows:
\begin{equation}
  \frac{D\tilde{\rho}}{D\tau} = -(\mathrm{div}\,{\bm \tilde{u}})\tilde{\rho},
\end{equation}
\begin{eqnarray}
  & & \frac{D(\mathrm{div}\,{\bm \tilde{u}})}{D\tau}+\frac{\partial \tilde{u}_j}{\partial
   \tilde{x}_i}\frac{\partial \tilde{u}_i}{\partial \tilde{x}_j}= \nonumber \\
  & & \left( \frac{1}{\tilde{\rho}}\frac{\partial}{\partial \tilde{x}_i}
 - \frac{1}{\tilde{\rho}^2}\frac{\partial \tilde{\rho}}{\partial
   \tilde{x}_i}\right) \left( \frac{\gamma-1}{\gamma}\frac{Pr}{Ep}
   \frac{\partial \tilde{\sigma}_{ij}}{\partial \tilde{x}_j}-\frac{\partial \tilde{P}}{\partial
    \tilde{x}_i}\right). 
\end{eqnarray}
These equations suggest
that the balance between the viscosity and the pressure gradient 
plays a crucial role in
the relaxation of the multidimensional localized
structures through the divergence of velocity.
Therefore, the viscosity should be considered as well as the pressure gradient
regardless of the dimension of the system.

\section{Discussion and concluding remarks}
\label{sec:discussion}

\begin{figure}
\begin{center}
\includegraphics[scale=0.65]{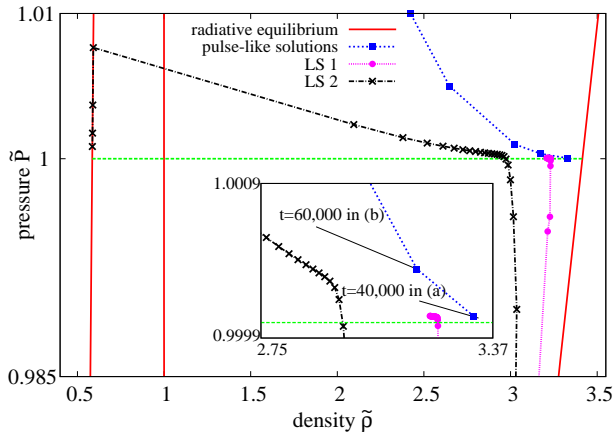}
\end{center}
\caption{
(Color online) Evolutions of peak pressure and peak density of the localized
 structures LS1 and LS2 shown in Fig.\ \ref{fig:peakdensity}.
The equilibrium curve and the pulselike stationary solution illustrated
in Fig.\ \ref{fig:tpgraph_nondim_1d} are also drawn.
The points on the evolution curves are plotted every $5000$ from $\tau=5000$
to $\tau =165 000$ for LS1 and from $\tau=5000$ to $\tau=135 000$ for LS2.
Inset: the evolutions of peak pressure and peak density magnified
around the times in which the localized structures are trapped near
the corresponding pulselike stationary solutions.
}
\label{fig:PSinP-rho}
\end{figure}

We have investigated localized cold structures in thermally unstable two-phase fluids.
The two-phase separation is induced by thermal instability 
via local heating and cooling
described by the Ginzburg-Landau-type heating-cooling function.
We have numerically obtained pulselike stationary solutions of which the peak densities
are not in the states of the radiative equilibrium 
in one-dimensional, two-dimensional axisymmetric,
and three-dimensional spherically symmetric systems using a shooting method.
Our solutions differ from the solutions obtained in the previous studies 
which consider bistable media and fronts connecting the two stable states
\cite{Elphick:1992,Aranson:1993,Shaviv:1994}.
The size of the solutions depends only on pressure, and the relationship 
between the size and the pressure qualitatively coincides with the results of
Nagashima \textit{et al.} \cite{Nagashima:2006}.

We have carried out direct numerical simulations of the thermally
unstable fluids with a small random density perturbation using a
high-accuracy spectral method.
During the two-phase separation, many cold clouds are formed,
and some of them merge or evaporate.
Finally, several cold clouds are left in the warm media and
persist for a long time.
The pulselike stationary solutions obtained by the shooting
 method are actually observed in the direct numerical simulations.
Such small and long-lived clouds have been observed in interstellar media 
and might be related with the long-lived localized clouds we have found
\cite{Nagashima:2006,Braun:2005,Stanimirovic:2005}.

We have shown that the viscosity balances with the pressure gradient in
the quasistationary state at higher order.
The balance remarkably suppresses the evaporation of the localized clouds.
Note that the viscous and the pressure-gradient effects are very weak
but play a crucial role in the persistence of the structures.
In contrast, most of the previous studies have neglected the viscous effect
(e.g., Refs.\ \cite{Elphick:1992,Aranson:1993,Shaviv:1994,Nagashima:2006,Inoue:2006}).
Our results strongly suggest that the viscosity changes the relaxation process
and suppresses the evaporation of the cold clouds.
We, therefore, expect that multidimensional clouds,
which are also pulselike as the one-dimensional clouds,
have a long relaxation time because of the viscosity.

Although the pulselike stationary solutions are obtained under
constant pressure, the pressure is a dependent variable in direct
numerical simulations. However, the balance between the viscosity
and the pressure gradient tends to keep the pressure nearly constant (pressure saturation).
In this situation, one of the pulselike stationary solutions is selected and attracts
the localized structure. This solution behaves similar to a saddlelike fixed
point in phase space. Figure \ref{fig:PSinP-rho} shows the evolution of
peak values of the localized structures LS1 and LS2 in density-pressure space.
The peak pressure and density of the
localized structure approach those of the pulselike stationary solution.
After a long stay around there, it quickly leaves from there to the lower equilibrium state.
Note that the leaving process of LS1 is not seen in Fig.\ \ref{fig:PSinP-rho}.
In this final escaping process, the profile of the localized structure
quickly deviates from the pulselike stationary solution with the same pressure value
as the peak pressure of the localized structure (see  Fig.\ \ref{fig:denssuccess}).
By this picture of the relaxation process, the relaxation time might be
dependent on the approaching process along with a stable manifold
of the pulselike stationary solution.
When similar to the case of LS2, the pressure saturation is not
sufficient, that is, localized structures are not close enough
to the stable manifold of one of the pulselike stationary solutions,
the localized structures seem to approach
a part of a manifold formed by the pulselike stationary solutions
rather than a single one, i.e., a fixed point.
Anyway, the dynamical system approach will help us
understand the formation process of localized clouds.

In the future, we plan to perform direct numerical simulations 
in two-dimensional axisymmetric and three-dimensional
spherically symmetric systems
in order to confirm the persistence of the multidimensional 
localized structures with the viscosity.
In addition, we will employ two-dimensional full numerical simulations
in order to investigate multidimensional effects such as interface dynamics.

\textbf{Acknowledgement:}
We would like to acknowledge stimulating discussions with Tsuyoshi Inoue and Shu-ichiro Inutsuka.
This work was supported by the Grant-in-Aid for the Global COE Program
"The Next Generation of Physics, Spun from Universality and Emergence"
from the Ministry of Education, Culture, Sports, Science and Technology (MEXT) of Japan.
The numerical calculations were carried out on SX8 at YITP in Kyoto University 
and scholar-processors at RIMS in Kyoto University.
S.T. was partly supported by the Grant-in-Aid for Scientific Research
 (C)(Grant No. 18540373) from the Japan Society for the Promotion of Science.


\begin{thebibliography}{10}

\bibitem{Toh:1987}
S.~Toh.
\newblock Statistical model with localized structures describing the
  spatio-temporal chaos of kuramoto-sivashinsky equation.
\newblock {\em J. Phys. Soc. Jpn}, 56:949, 1987.

\bibitem{Iwasaki_Toh:1992}
H.~Iwasaki and S.~Toh.
\newblock Statistics and structures of strong turbulence in a complex
  ginzburg-landau equation.
\newblock {\em Prog. Theor. Phys.}, 87:1127, 1992.

\bibitem{Itano_Toh:2001}
T.~Itano and S.~Toh.
\newblock The dynamics of bursting process in wall turbulence.
\newblock {\em J. Phys. Soc. Jpn}, 61:703, 2001.

\bibitem{Kawahara_Kida:2001}
G.~Kawahara and S.~Kida.
\newblock Periodic motion embedded in plane couette turbulence: regeneration
  cycle and burst.
\newblock {\em J.~Fluid Mech.}, 449:291, 2001.

\bibitem{Pipe:2007}
B.~Eckhardt, T.~M. Schneider, B.~Hof, and J.~Westerweel.
\newblock Turbulence transition in pipe flow.
\newblock {\em Ann. Rev. Fluid Mech.}, 39:447, 2007.

\bibitem{Waleffe:2003}
F.~Waleffe.
\newblock Homotopy of exact coherent structures in plane shear flows.
\newblock {\em Phys. Fluids}, 15:1517, 2003.

\bibitem{Davidson:1972}
K.~Davidson.
\newblock Photoionization and the emission-line spectra of quasi-stellar
  objects.
\newblock {\em Astrophys. J.}, 171:213, 1972.

\bibitem{Lipschultz:1987}
B.~Lipschultz.
\newblock Review of marfe phenomenas in tokamaks.
\newblock {\em J. Nucl. Mater.}, 145-147:15, 1987.

\bibitem{Meerson:1996}
B.~Meerson.
\newblock Nonlinear dynamics of radiative condensations in optically thin
  plasmas.
\newblock {\em Rev. Mod. Phys.}, 68:215, 1996.

\bibitem{Audit:2005}
E.~Audit and P.~Hennebelle.
\newblock Thermal condensation in a turbulent atomic hydrogen flow.
\newblock {\em Astron. Astrophys.}, 433:1, 2005.

\bibitem{Heitsch:2005}
F.~Heitsch, A.~Burkert, L.~W. Hartmann, A.~D. Slyz, and J.~E.~G. Devriendt.
\newblock Formation of structure in molecular clouds: A case study.
\newblock {\em Astrophys. J. Lett.}, 633:L113, 2005.

\bibitem{Vazquez:2006}
E.~V\'azquez-Semadeni, D.~Ryu, T.~Passot, R.~F. Gonz\'alez, and A.~Gazol.
\newblock Molecular cloud evolution. {I}. molecular cloud and thin cold neutral
  medium sheet formation.
\newblock {\em Astrophys. J.}, 643:245, 2006.

\bibitem{Elphick:1992}
C.~Elphick, O.~Regev, and N.~Shaviv.
\newblock Dynamics of fronts in thermally bistable fluids.
\newblock {\em Astrophys. J.}, 392:106, 1992.

\bibitem{Aranson:1993}
I.~Aranson, B.~Meerson, and P.~V. Sasorov.
\newblock Nonlinear radiative-condensation instability and pattern formation:
  One-dimensional dynamics.
\newblock {\em Phys. Rev. E}, 47:4337, 1993.

\bibitem{Braun:2005}
R.~Braun and N.~Kanekar.
\newblock Tiny h i clouds in the local ism.
\newblock {\em Astron. Astrophys. Lett.}, 436:L53, 2005.

\bibitem{Stanimirovic:2005}
S.~Stanimirovi\'c and C.~Heiles.
\newblock The thinnest cold h i clouds in the diffuse interstellar medium?
\newblock {\em Astrophys. J.}, 631:371, 2005.

\bibitem{Nagashima:2006}
M.~Nagashima, S.~Inutsuka, and H.~Koyama.
\newblock How long can tiny hi clouds survive ?
\newblock {\em Astrophys. J. Lett.}, 652:L41, 2006.

\bibitem{Parker:1953}
E.~N. Parker.
\newblock Instability of thermal fields.
\newblock {\em Astrophys. J.}, 117:431, 1953.

\bibitem{Field:1969}
G.~B. Field, D.~W. Goldsmith, and H.~J. Habing.
\newblock Cosmic-ray heating of the interstellar gas.
\newblock {\em Astrophys. J. Lett.}, 155:L149, 1969.

\bibitem{Field:1965}
G.~B. Field.
\newblock Thermal instability.
\newblock {\em Astrophys. J.}, 142:531, 1965.

\bibitem{Zeldovich:1969}
Y.~B. Zel'dovich and S.~B. Pikel'ner.
\newblock The phase equilibrium and dynamics of a gas volume that is heated and
  cooled.
\newblock {\em Sov. Phys. JETP}, 29:170, 1969.

\bibitem{Graham:1973}
R.~Graham and W.~D. Langer.
\newblock Pressure equilibrium of finite-size clouds in the interstellar
  medium.
\newblock {\em Astrophys. J.}, 179:469, 1973.

\bibitem{Elphick:1991}
C.~Elphick, O.~Regev, and E.~A. Spiegel.
\newblock Complexity from thermal instability.
\newblock {\em Mon. Not. R. Astron. Soc.}, 250:617, 1991.

\bibitem{Nagashima:2005}
M.~Nagashima, H.~Koyama, and S.~Inutsuka.
\newblock Evaporation and condensation of hi clouds in thermally bistable
  interstellar media: semi-analytic description of isobaric dynamics of curved
  interfaces.
\newblock {\em Mon. Not. R. Astron. Soc. Lett.}, 361:L25, 2005.

\bibitem{Koyama:2000}
H.~Koyama and S.~Inutsuka.
\newblock Molecular cloud formation in shock-compressed layers.
\newblock {\em Astrophys. J.}, 532:980, 2000.

\bibitem{Koyama:2002}
H.~Koyama and S.~Inutsuka.
\newblock An origin of supersonic motions in interstellar clouds.
\newblock {\em Astrophys. J. Lett.}, 564:L97, 2002.

\bibitem{Shaviv:1994}
N.~J. Shaviv and O.~Regev.
\newblock Interface dynamics and domain growth in thermally bistable fluids.
\newblock {\em Phys. Rev. E}, 50:2048, 1994.

\bibitem{Inoue:2006}
T.~Inoue, S.~Inutsuka, and H.~Koyama.
\newblock Structure and stability of phase transition layers in the
  interstellar medium.
\newblock {\em Astrophys. J.}, 652:1331, 2006.

\end{thebibliography}
\end{document}